%% file: main.tex
\documentclass[conference]{IEEEtran}

\usepackage{cite}
\usepackage{amsmath,amssymb,amsfonts}
\usepackage{algorithmic}
\usepackage{graphicx}
\usepackage{textcomp}
\usepackage{xcolor}
\usepackage{hyperref}
\usepackage{tikz}
\usepackage{listings,fancyvrb}
\usepackage{tabularx}
\usepackage{verbatim}
\usepackage[T1]{fontenc}

\usepackage{caption}
\usepackage{xspace}
\newcommand{\pimod}{PIMOD\xspace}

\input{_orcid}

\lstdefinestyle{common}{
    xleftmargin=1.5em,
    xrightmargin=.5em,
    frame=single,
    framesep=.5em,
    framerule=0pt,
    fancyvrb=true,
    numbers=left,
    numberstyle=\scriptsize,
    basicstyle=\ttfamily,
    keywordstyle=\color{cyan!50!blue!75!black}\bfseries,
    commentstyle=\color{red!50!black}\itshape,
    stringstyle=\ttfamily\color{green!50!black},
    emphstyle=\color{red},
    backgroundcolor=\color{blue!5},
    showspaces=false,
    showstringspaces=false,
    fontadjust=true,
    keepspaces=true,
    flexiblecolumns=true,
    captionpos=b,
    breaklines=true,
}

\lstdefinestyle{Pimod}{
    style=common,
    language=bash,
    morestring=[d]{EOF},
    otherkeywords = {FROM, TO, INPLACE, PUMP, RUN, HOST, INSTALL, \$DEVICE, \$HOME, <, >, GOOS=linux, GOARCH=arm, GOARM=5},
    morekeywords = [2]{\$DEVICE, \$HOME, GOOS=linux, GOARCH=arm, GOARM=5},
    keywordstyle = [2]{\color{green!50!black}},
    aboveskip=10pt,
}
\lstdefinestyle{console}{
    style=common,
    basicstyle=\footnotesize,
    language=bash,
    morestring=[d]{EOF},
    otherkeywords = {},
    aboveskip=10pt,
}

\begin{document}

\def\arraystretch{1.3}

\title{\pimod: A Tool for Configuring Single-Board Computer Operating System Images}

\author{
\IEEEauthorblockN{
    Jonas Höchst\IEEEauthorrefmark{1}\IEEEauthorrefmark{2} \orcidicon{0000-0002-7326-2250}, 
    Alvar Penning\IEEEauthorrefmark{1} \orcidicon{0000-0003-2992-1316}, 
    Patrick Lampe\IEEEauthorrefmark{1}\IEEEauthorrefmark{2}\orcidicon{0000-0002-6233-0959}, 
    Bernd Freisleben\IEEEauthorrefmark{1}\IEEEauthorrefmark{2} \orcidicon{0000-0002-7205-8389}
}
\IEEEauthorblockA{
    \\
    \IEEEauthorrefmark{1} Department of Mathematics \& Computer Science, University of Marburg, Germany\\
    \{%
        \href{mailto:hoechst@informatik.uni-marburg.de}{hoechst}, 
        \href{mailto:penning@informatik.uni-marburg.de}{penning}, 
        \href{mailto:lampep@informatik.uni-marburg.de}{lampep}, 
        \href{mailto:freisleb@informatik.uni-marburg.de}{freisleb}%
    \}@informatik.uni-marburg.de\\
    \IEEEauthorrefmark{2} Department of Computer Science / Electrical Engineering \& Information Technology, TU Darmstadt, Germany\\
    \{%
        \href{mailto:jonas.hoechst@maki.tu-darmstadt.de}{jonas.hoechst}, 
        \href{mailto:patrick.lampe@maki.tu-darmstadt.de}{patrick.lampe}, 
        \href{mailto:bernd.freisleben@maki.tu-darmstadt.de}{bernd.freisleben}%
    \}@maki.tu-darmstadt.de
}
}

\maketitle

\begin{abstract}
\input{0_abstract}
\end{abstract}

\begin{IEEEkeywords}
\input{0_keywords}
\end{IEEEkeywords}

\input{1_intro}

\input{2_related_work}
\input{3_design}
\input{4_impl}
\input{5_eval}
\input{6_conclusion}

\bibliographystyle{IEEEtranS}
\bibliography{main}

\end{document}

%% file: _orcid.tex
\usepackage{tikz}
\usepackage{scalerel}
\usetikzlibrary{svg.path}

\definecolor{orcidlogocol}{HTML}{A6CE39}
\tikzset{
  orcidlogo/.pic={
    \fill[orcidlogocol] svg{M256,128c0,70.7-57.3,128-128,128C57.3,256,0,198.7,0,128C0,57.3,57.3,0,128,0C198.7,0,256,57.3,256,128z};
    \fill[white] svg{M86.3,186.2H70.9V79.1h15.4v48.4V186.2z}
    svg{M108.9,79.1h41.6c39.6,0,57,28.3,57,53.6c0,27.5-21.5,53.6-56.8,53.6h-41.8V79.1z M124.3,172.4h24.5c34.9,0,42.9-26.5,42.9-39.7c0-21.5-13.7-39.7-43.7-39.7h-23.7V172.4z}
    svg{M88.7,56.8c0,5.5-4.5,10.1-10.1,10.1c-5.6,0-10.1-4.6-10.1-10.1c0-5.6,4.5-10.1,10.1-10.1C84.2,46.7,88.7,51.3,88.7,56.8z};
  }
}

\newcommand\orcidicon[1]{\href{https://orcid.org/#1}{\mbox{\scalerel*{
\begin{tikzpicture}[yscale=-1,transform shape]
\pic{orcidlogo};
\end{tikzpicture}
}{|}}}}


%% file: 0_abstract.tex
Computer systems used in the field of humanitarian technology are often based on general-purpose single-board computers, such as Raspberry Pis. 
While these systems offer great flexibility for developers and users, configuration and deployment either introduces overhead by executing scripts on multiple devices or requires deeper technical understanding when building operating system images for such small computers from scratch.
In this paper, we present \pimod, a software tool for configuring operating system images for single-board computer systems.
We propose a simple yet comprehensive configuration language. 
In a configuration profile, called Pifile, a small set of commands is used to describe the configuration of an operating system image.
Virtualization techniques are used during the execution of the profile in order to be distribution and platform independent. 
Commands can be issued in the guest operating system, providing access to the distribution specific tools, e.g., to configure hardware parameters.
The implementation of \pimod is made public under a free and open source license.
\pimod is evaluated in terms of user benefits, performance compared to on-system configuration, and applicability across different hardware platforms and operating systems.

%% file: 0_keywords.tex
Single-Board Computer, 
Operating System Image, 
System Provisioning

%% file: 1_intro.tex
\section{Introduction}
\label{sec:intro}

When applying technology in the humanitarian field, it is particularly important that the equipment used is available and that the installations are traceable and maintainable by the user groups.
For this reason and for reasons of low cost, single-board computers (SBCs), such as the Raspberry Pi, are often used as the basis for research and especially for practical applications.
Various use cases have been posted, where such devices are key enablers for the proposed solutions, be it technical or general education \cite{srinivasan2013greeneducomp, yamanoor2017high},
monitoring of technology \cite{truitt2019low} or monitoring in the health care sector \cite{kumar2016iot}, or various communication technologies \cite{quitevis2018feasibility, baumgaertner2019smart, baumgartner2018environmental}.

When dealing with single-board computers, either for software development or when deploying hardware based on these boards, there is a lack of support for creating operating system (OS) images.
There are several cases in which readily configured images and use case specific distributions need to be distributed to users or operators. 
Devices like a Raspberry Pi are used at home, in applications for multimedia centers or smart homes, but also in challenging applications such as emergency response, environmental monitoring, Internet-of-Things (IoT), and smart city infrastructures.

Single-board platforms that do rely on an operating system regularly use images provided by vendors or third parties. 
Typically, an image is flashed to an SD card and then booted in a system.
Since there is no installation process, the OSes heavily depend on defaults, e.g., username, password, installed software, or on scripts executed on the first boot, e.g., cryptographic parameters or partition size adjustments.
Software can then be installed and configurations can be adapted in the running system. 
While this seems to be convenient for single deployments and fast progress compared classical installation routines, it is not suitable for larger deployments. 

When custom software and additional configurations need to be added to an OS image, this can either be achieved by 
a) creating an OS image from scratch, 
b) adding scripts to be run on the first boot, or
c) create an image from a previously configured system.
However, these methods each have their individual drawbacks.
Bootstrapping images from scratch requires deep technical understanding. When using first boot scripts, a network connection is required on this first boot. Creating an image from a configured system requires additional steps to revert OS specific first boot configurations. 

In this paper, we present \pimod, a tool for modifying an existing operating system image by executing commands described in a configuration file. 
In the proposed line-based configuration, a \texttt{Pifile}, i.e., a small set of commands, can be used to describe how an image will be created.
These commands are then interpreted by \pimod and executed accordingly.
In our approach, the target image is based on an existing image, it then can be resized, and files from the host system can be included in the image.
The special \texttt{RUN} command allows running commands inside the image, so that guest OS specific packet managers and configuration tools can be used.
Our approach can easily be used with continuous integration (CI) systems and enable reproducible builds of single-board computer operating system images.
The software as described in this paper is released under the free and open source GPL-3.0 license\footnote{\url{https://www.gnu.org/licenses/gpl-3.0.html}} and is available online\footnote{\url{https://github.com/nature40/pimod/}}.

To summarize, we make the following contributions:
\begin{itemize}
    \item We propose a novel method of configuring single-board computer operating system images.
    \item We present a simple yet comprehensive operating system image configuration language.
    \item We provide a free and open source implementation of \pimod. 
    \item We conduct an evaluation of \pimod in terms of user benefits, performance, and language flexibility. 
\end{itemize}

The paper is organized as follows.
Section~\ref{sec:relwork} discusses related work. 
In Section \ref{sec:design}, we present requirements and designs decisions.
Section~\ref{sec:impl} discusses implementation issues.
Section \ref{sec:eval} presents experimental results.
Section \ref{sec:conclusion} concludes the paper and outlines areas of future work.

%% file: 2_related_work.tex
\section{Related Work}
\label{sec:relwork}

In general, single-board computers (SBCs) use operating system images provided by 
a) the hardware vendor, 
b) a third party operating system distributor, or
c) a software provider, bundling up its software and dependencies to create a software-specific distribution.
The images are flashed to a SD card and then booted on a SBC.
Since there is no installation process, the OSes heavily depend on defaults and can only be adapted by running them and changing software or configurations. 

As a first option, use case specific images can be created by using the tools provided by the SBC vendor, such as pi-gen provided by the Raspberry Pi Foundation\cite{pigen} or alternative approaches \cite{Kairiukstis2012BuildRaspbianImage}.
The tool is designed to create images from scratch and highly adjusted to the specific use case.
The open source wireless router distribution OpenWRT features its own build system \cite{fainelli2008openwrt}.
This build system is created modularly, and own packages can easily be integrated into the build process.
In addition to build images from scratch, the authors created an image builder, specifically targeted for OpenWRT, which installs precompiled packages to an image.
However, the image builder is targeted specifically for their operating system and does not work for others.
These tools often result in long execution times, since all components are installed or even cross-compiled from scratch. 

The second available option consists of tools that add custom scripts to be executed on the devices itself.
With PiBakery, a graphical configuration interface for Raspbian is provided, which then creates scripts that are executed on the first boot or on every boot accordingly \cite{PiBakery}.
Some distributions use the first boot for configuration, e.g., ssh keys in the case of Raspbian, which need to be taken care of manually \cite{raspbian}.
These tools have the drawback that, e.g., requested software needs to be installed on every device independently, which results in multiple identical installation processes that may lead to high network overheads. 
The approach also lacks the possibility of being integrated into Continuous Integration (CI) build systems.

The third widely available option is to use an existing SD card with installed software and a finished configuration.
While this is a straightforward approach, it can hardly be automated. 
To be storage efficient, the copied image, including the partition table and file system, would need to be shrinked, which requires additional tools, such as PiShrink \cite{pishrink}.

When dealing with configuration of systems, Docker is a well known virtualization system, designed for dependency management and containerization of applications \cite{merkel2014docker}.
Docker features a simple imperative configuration language.
A new image is built based on an old image and extended by copying files, altering a Docker specific configuration or running commands inside the container.
When using Docker to provide and install software, it is necessary to install Docker and the corresponding software images on the live system.
Therefore, this approach does not overcome the problem of multiple installations on individual devices and does not offer a full operating system image. 

For configuration management tools like Ansible \cite{hochstein2017ansible}, Saltstack \cite{hosmer2012getting}, and Puppet \cite{loope2011managing}, the main concept is to have a central server that ships a configuration to every node.
The node then adapts the installed system in the manner defined in the configuration. 
This method has the drawback that it uses more network resources because every single node has to download updates and the installer for itself.  
Also, the nodes have to be booted so that the first boot scripts are executed. 
The possibility of configuring and reconfiguring a running system is quite helpful, but we focus on the creation of full OS system images with preinstalled and configured software. 
Furthermore, the client part has to be installed on every single node, and the master node has to run when a new node should be configured. 

%% file: 3_design.tex
\section{\pimod Design}
\label{sec:design}

The goal of \pimod is to facilitate the creation of single-board computer operating system images with custom software in an easy and reproducible manner and simplify the deployment of such devices.
To reach this goal, a simple yet comprehensive configuration language is provided, which is interpreted to modify a system image.
The language should be manageable through versioning systems to support the overall goal of reproducibility.
A generic configuration language cannot rely on distribution-specific configuration parameters and thus should provide an interface to the distribution's configuration mechanisms.
With \pimod we target Linux-based operating systems, which are widespread in several communities using SBCs \cite{baun2016mobile, johnston2018commodity}.

\subsection{The \pimod Language}
In this section, the \pimod language used in a Pifile is presented.
To reach the goal of easy learnability, the language was inspired by the Dockerfile language, which is already widely known.
A Pifile is a line-based document where each line may either contain 
a) an empty line that may contain white space,
b) a comment indicated by a hash symbol,
c) a \pimod command written in caps followed by parameters.

\paragraph{FROM <source> [partition]} The required source parameter declares a base image to be found in the local file system, a block device to create an image from, or an URL to be downloaded and extracted.
Optionally, the partition number resized and mounted in the further process can be declared. It defaults to the second partition, since most operating systems use one boot as one system partition.

\paragraph{TO <destination image>} When a Pifile is executed, the resulting image is written next to the Pifile and named after the respective Pifile. 
The image destination can be changed by running the \texttt{TO} command. 
When a block device is specified, the defined source is written to the respective device and further commands are executed directly on the device.

\paragraph{INPLACE <image>} Using the \texttt{INPLACE} command, an image can be specified on which the commands are executed.

\paragraph{PUMP <bytes>} Using the \texttt{PUMP} command, the image is increased by the given amount of bytes, SI prefixes such as k, M, G or, T are supported.

\paragraph{PATH <location>} By default, the local \texttt{PATH} variable of the host system is used inside the guest system.
With this command, it can be extended by another location.

\paragraph{RUN <cmd>} Commands specified using the \texttt{RUN} command are executed inside of the operating system image.
Note that the operating system of the image is not started, but the run time environment of the target system is modeled.

\paragraph{INSTALL [mode] <source> <destination>} Installing custom files from the host system is especially useful when custom software is used, or for configuration purposes.
The source parameter relates to a file in the host file system, the destination describes a path in the file system of the target system.
The optional mode parameter can be useful when installing executables, e.g., cross-compiled software.

\paragraph{HOST <cmd>} When a command is specified using the HOST command, it is executed on the local system rather than inside the image.
Issuing a local command can especially be useful for preparing configuration files or cross-compiling software, which later is installed to the guest system.

\begin{lstlisting}[
    style=Pimod, 
    caption={\pimod example 1: upgrade Raspbian and enable the serial console.},
    label={lst:example},
    float=tp,
]
FROM 2020-05-27-raspios-buster.img 2
TO raspbian-buster-upgraded.img

# Increase the image by 100 MB
PUMP 100M

# Enable serial console using built-in configuration tool
RUN raspi-config nonint do_serial 0

# Upgrade the operating system image
RUN apt-get update
RUN bash -c 'DEBIAN_FRONTEND=noninteractive apt-get -y dist-upgrade'

# Install an ssh key
INSTALL id_rsa.pub /home/pi/.ssh/authorized_keys
\end{lstlisting}

In Listing~\ref{lst:example}, a Pifile is presented that features all commands of the \pimod language. 
Line 1 defines a source image to be found in the local file system and the partition to be resized and mounted as the primary system partition.
In Line 2, we declare that the file should be written to an alternative location.
Line 5, \texttt{PUMP 100M}, causes the image and the second partition to be increased by 100 mebibytes.
In Line 8, a distribution-specific configuration tool is used to enable the serial console available at the target hardware.
Line 11 and 12 are used to upgrade the operating system by first updating the sources of the packet manager and then running a distribution update.
Note that in Line 12 an environment variable is set by running the command inside a bash shell.
Finally, in Line 15, a ssh public key is copied to allow remote login.

\subsection{Linux Support}
The Pifile language is designed to be a simple yet comprehensive operating system image configuration language. 
To reach this goal, some assumptions were made during the design phase. 
First, to enable fast execution of Pifiles, we do not want to use full system emulation, which would result in booting the guest system kernel.
This would have the disadvantage that, e.g., the first boot scripts of the distribution would be executed and other parameters would be initialized, such as cryptographic keys, as discussed in the introduction.
We decided to use a QEMU-based system emulation, which allows us to execute Linux ELF binaries across multiple different instruction set architectures \cite{bellard2005qemu}. 
Second, especially mounting the partitions of the image according to the distribution requires specific knowledge, which is hard to generalize.
Therefore, we decided to use the file system table defined by the Filesystem Hierarchy Standard, \texttt{/etc/fstab}, which itself is used by many Linux operating systems.
Third, the executed binaries are searched according to a path variable, which itself is distribution specific.
In \pimod, this variable is initialized from the host system and can be extended by using the \texttt{PATH} command in a Pifile.

\subsection{Continuous Integration Support}
Continuous Integration (CI) is a technique used to overcome integration problems in the development cycle during software engineering.
It has been shown that continuous integration improves the productivity of project teams and boosts the integration of external contributions without a reduction in code quality \cite{vasilescu2015quality}.
When combining version management and modern CI systems, every commit of a software under development is automatically integrated into a larger context and tested.
\pimod is designed to be used in combination with CI to create software-specific operating system images in a reproducible and easy manner.

\subsection{Host System Support} 
Another goal of our approach is extensibility, such that it can be integrated into workflows of the communities using \pimod.
Thus, the configuration language should provide options to interface the host system.
One option to enable interfacing in this manner is the already described \texttt{HOST} command. 
In addition, users should be able to use environment variables defined in the host system and program a control flow.

%% file: 4_impl.tex
\section{Implementation}
\label{sec:impl}

The target of modifying system images and executing code inside of a system image can be achieved best by using system tools.
GNU/Linux ships several helpful tools for the individual tasks implemented by \pimod.
To make use of and integrate existing tools in a simple manner, \pimod was implemented using the Bash programming language\cite{ramey1994s}.

\begin{figure}[tb]
    \centering
    \includegraphics[width=\columnwidth]{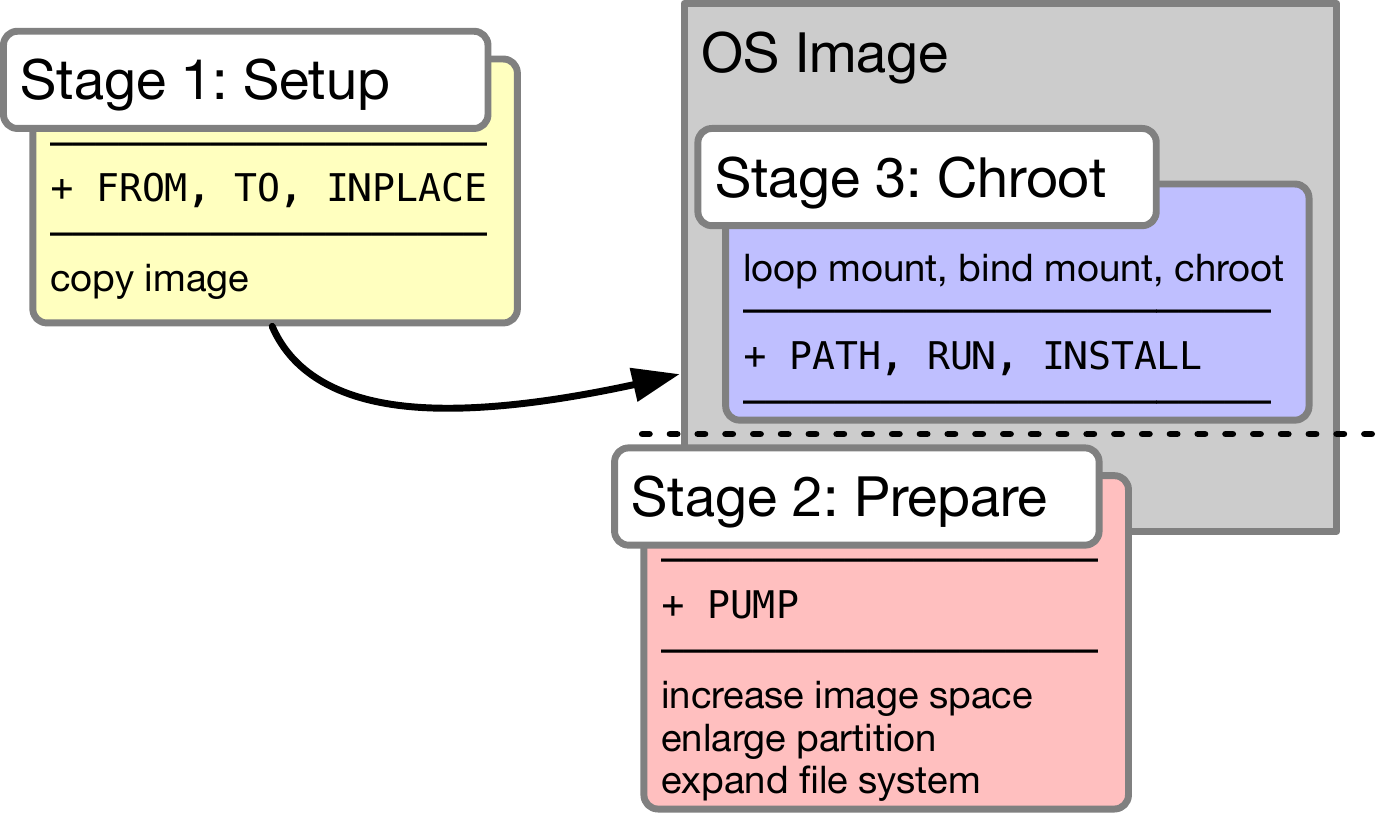}
    \caption{Stages of \pimod: preparation, commands, and post-processing.}
    \label{figure:pimod_stages}
\end{figure}

The interpretation of a Pifile is implemented in several stages, since some commands can only be executed after others, as shown in Fig.~\ref{figure:pimod_stages}. 
We define three stages to which the commands are assigned:
First, \textit{setup}, managing the \texttt{FROM}, \texttt{TO} and \texttt{INPLACE} configuration.
Second, \textit{prepare}, handling changes of the image itself, currently only depicting \texttt{PUMP}.
Third, \textit{chroot}, covering all commands interacting with the file system of an image, namely \texttt{RUN}, \texttt{PATH} and \texttt{INSTALL}.
During the execution of a Pifile, it is actually executed one time for each stage, only executing the commands belonging to the individual stage. 
This mechanism insures that source and target are defined before resizing an image, which itself needs to happen before modifying content on the guest.  

\subsection{Stage 1: Setup}
In the setup stage, the \texttt{FROM}, \texttt{TO}, and \texttt{INPLACE} commands are executed, checking and setting source and destination system images.  
\texttt{FROM} is not only able to handle local images, but can also download images from a remote location by specifying an HTTP(s) or FTP URL. 
A local cache minimizes network load and timely overhead.
When no destination is defined via \texttt{TO}, it is derived from the name of the Pifile.
The stage is concluded by copying the source image to the destination location.
In the special case of an identical source and destination path and when using \texttt{INPLACE}, the modifications are executed in place.

\subsection{Stage 2: Prepare}
The stage implements the \texttt{PUMP} command.
Enlarging an image requires to 
a) increase the image file size,
b) enlarge a partition inside the image, and
c) expand the file systems to the size defined by the partition.
These subtasks are implemented using the GNU/Linux utilities
\texttt{dd}\footnote{\url{https://www.gnu.org/software/coreutils/manual/html\_node/dd-invocation.html}}, 
\texttt{fdisk}\footnote{\url{https://www.gnu.org/software/fdisk/}}, and 
\texttt{resize2fs}\footnote{\url{https://linux.die.net/man/8/resize2fs}}.

\subsection{Stage 3: Chroot}
Before executing the \texttt{RUN} and \texttt{INSTALL} commands implemented by this stage, some preparations need to be taken:
first, the system image file is associated with a loop device of the host system.
Then, the main partition's file system is mounted inside the host system. 
A working chroot environment requires system interaction, which can easily achieved by importing \texttt{/dev}, \texttt{/sys}, \texttt{/proc} and \texttt{/dev/pts} using a bind mount.
The network interfaces are available through the host system kernel, the domain name system configuration is done by bind mounting \texttt{/etc/resolv.conf}.
After this step, the statically linked QEMU binaries for the supported platforms are also bind mounted in the chroot environment.
Ultimately, additional partitions defined in the file system table of the guest system are mounted.
\texttt{INSTALL} is implemented by copying the requested files to the target file system and optionally adjusting the permissions. 
Running a command inside the target image is easily achievable using chroot:
a command is executed in a specified root directory and thus using all binaries, libraries and resources of the mounted image.

\subsection{Continuous Integration}
\pimod is designed to work in combination with continuous integration services.
We provide an example integration for two different CI services.
Travis CI
is a free and open source CI service, which has been shown to be used by a wide variety of software projects \cite{beller2017oops}.
GitHub Actions
is a CI service integrated with GitHub, a software hosting platform, widely used for Open Source projects \cite{daigle2018github}.
In both integrations, first the dependencies need to be installed, then the resources, such as a base image are downloaded, and finally the Pifile is executed. 
The output of \pimod is presented inside the web interfaces of the individual service. 
Also, both implementations offer the possibility to release the created image in the form of a downloadable image.
Hence, the developers of a use case specific distribution can test their progress locally and tag a specific git commit.
This indeed triggers a cloud build using the discussed CI integration and uploads an image to the corresponding releases web page.
Our example integration is also available free and open source in a separate repository\footnote{\url{https://github.com/nature40/pimod-ci/}} and can easily be forked and adapted.

%% file: 5_eval.tex
\section{Evaluation}
\label{sec:eval}

In this section, we evaluate \pimod. 
First, the benefits of using \pimod from the view of a sensor network operator are discussed.
Then, the performance of the approach is evaluated by comparing execution times of exemplary commands.
Third, the generalization of \pimod is investigated by testing the software with Linux OS images of different distributions and made for different hardware.

\subsection{\pimod vs. Manual Integration}
The first goal of \pimod is to facilitate the modification of single-board computer operating system images and thereby simplify deployment.
We evaluate this goal by discussing the use case of deploying nodes in a sensor network scenario.
The deployment of such sensor nodes can be done in different ways.
A straightforward approach would be to repeat the deployment on each node manually.
First, a chosen operating system is installed, then the operator connects to each node, installs dependencies and software, and configures the system.
A more complex yet more efficient way would be to install only one system by the steps presented above and clone this installation to the other systems.
In some cases, the operator would need to manually alter some configurations, done during the first boot.
This, however, requires the appliance to be connected to a fast Internet connection and checks on every device that the initial scripts did run correctly.
The third alternative for the operator would be to build the system image by him- or herself and execute the required steps in the process. 
This would, however, require a deep understanding of the build process of an operating system, which can take time to understand and which itself takes a certain time to execute.
In addition, some software build systems require vast amounts of resources, e.g., TensorFlow, a machine learning toolkit. 
Building the software requires certain tricks, e.g., swap partitions to allow for larger amounts of memory, which can be circumvented using more powerful hardware.
With \pimod, the operators can create a configuration file in which all the required steps can be described.
The resulting Pifile can then be executed either locally resulting in an image to be flashed or integrated into a CI system, e.g., to build and upload an image to a certain online location.
The approach can also be used to write the resulting image to a SD card, e.g., to write device-specific configuration files.

\subsection{Performance Evaluation}

\begin{table*}[!t]
\centering
\begin{minipage}{\textwidth}
\caption{Example executions times of different commands using a Raspberry Pi compared to \pimod.}
\label{table:runtimes}
\begin{tabularx}{\textwidth}{l|Xrrr}
\textbf{Operating System} & \textbf{Command}                                        & $\mathbf{t_{RasPi}}$ & $\mathbf{t_{\pimod}}$ & \textbf{Overhead} \\ \hline
OpenWRT               & \texttt{opkg update}                                    &  2.86 s                   &  3.11 s           &   8.74 \%         \\
                      & \texttt{opkg install python}                            & 19.77 s                   & 32.53 s           &  64.54 \%         \\
                      & \texttt{wget -O /dev/null http://host/100m.bin}          &  8.92 s                   &  0.94 s           & -90.51 \%         \\
                      & \texttt{wget -O /dev/null https://host/100m.bin}         &  9.29 s                   & 10.55 s           &  13.56 \%         \\ \hline
Raspbian              & \texttt{apt-get update}                                 & 19.61 s                   & 14.61 s           & -25.49 \%         \\
                      & \texttt{apt-get install -y python3}                     & 60.73 s                   & 56.38 s           &  -7.16 \%         \\
                      & \texttt{dd if=/dev/urandom of=/dev/null}                &  2.63 s                   &  0.50 s           & -80.99 \%         \\
                      & \texttt{dd if=/dev/urandom of=100m.bin}                 &  7.19 s                   &  0.63 s           & -91.24 \%         \\
                      & \texttt{openssl enc -aes-256-cbc}                       &  5.52 s                   &  3.94 s           & -28.62 \%        
\end{tabularx}
\end{minipage}
\end{table*}
To evaluate the performance of \pimod, multiple example commands are executed on a single-board computer and using \pimod. 
The experiments presented in Table~\ref{table:runtimes} were executed on a Raspberry Pi 3 Model B V1.2 that consists of an ARM-Cortex-A53 with 4 cores of 1,2 GHz and 1 GB RAM.
For storage, a Samsung EVO Plus 32 GB microSD of Ultra High Speed (UHS) class U1 was used, which allows read speeds of 95 MB/s and write speeds of 20 MB/s.
The experiments of \pimod were conducted on a x86\_64-based server featuring two Intel Xeon E5-2698 CPUs, a total of 256 GB RAM. 
Storage was realized using a NVMe-based Intel SSDPEKKW512G7, with 512 GB storage and a read and write speed of up to 1775 MB/s and 560 MB/s. 
All tests were repeated 5 times and averaged.
Although the systems themselves are quite different in terms of performance, we try to mimic a build server that might be running in a continuous integration pipeline.

\begin{figure}[tb]
    \centering
    \vspace{.05in}
    \includegraphics[width=\columnwidth]{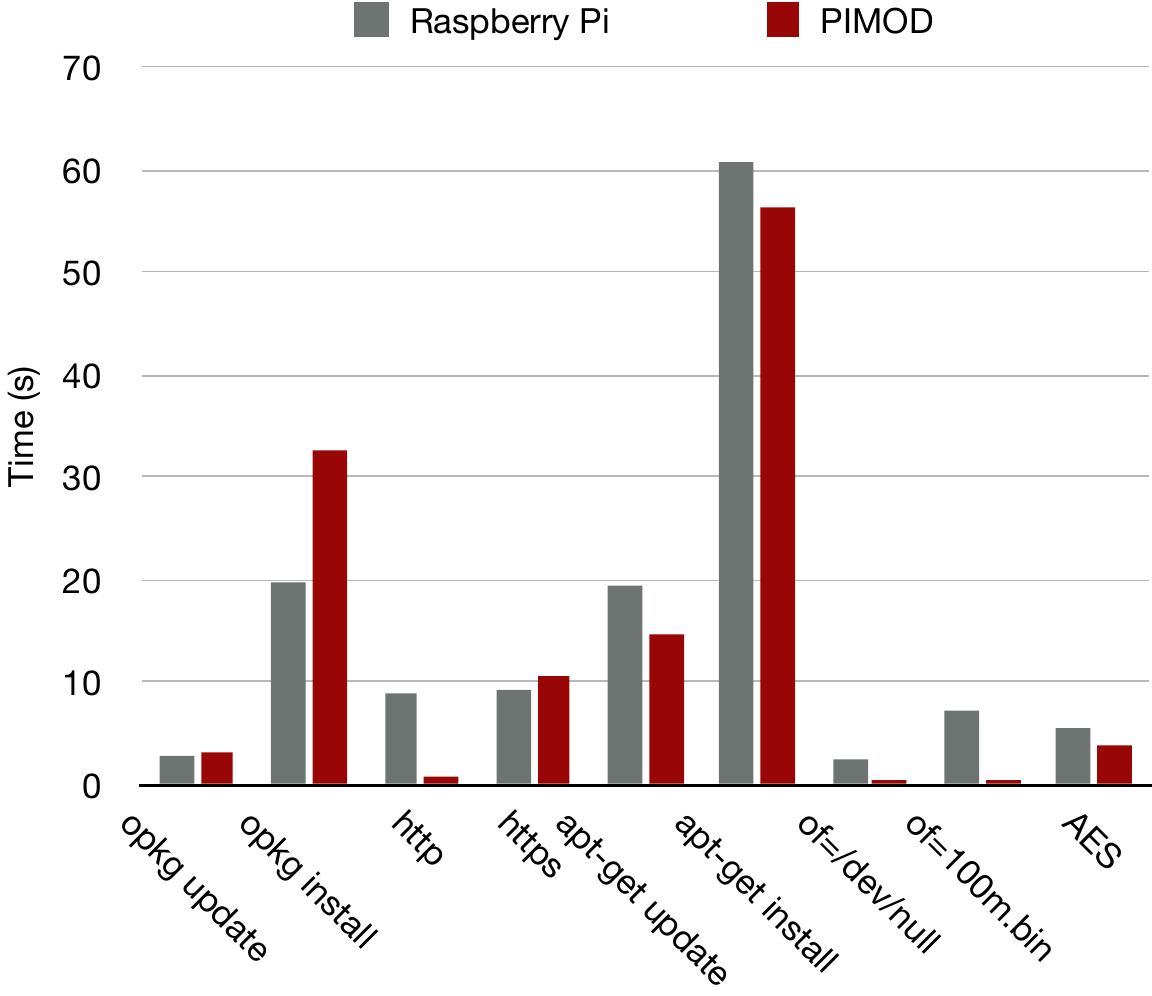}
    \caption{Example executions times of different commands using a Raspberry Pi compared to \pimod.}
    \label{figure:runtimes}
\end{figure}

In Table~\ref{table:runtimes}, the runtimes of various commands on different operating systems are presented, namely OpenWRT and Raspbian Buster.
With \texttt{opkg update} and \texttt{apt-get update}, the individual packet managers update the list of available packages. 
The OpenWRT packet manager introduces a small overhead using \pimod compared to native execution.
Updating the package list consists of downloading the compressed lists and verifying the signature, especially the latter requires many ARM instructions to be simulated on the x86\_64 hardware and therefore introduces some overhead. 
Installing software using \texttt{opkg} is faster on the native hardware compared to \pimod, which is mostly related to the used \texttt{gzip} compression executed through \texttt{qemu}.

When looking at Raspbian Buster's packet manager \texttt{apt}, \pimod achieves better runtimes compared to native execution, for both updating the package.
One reason for this is the much faster write speed of the desktop system compared to the microSD card of the Raspberry Pi.
Since OpenWRT is designed for low footprint systems, such as routers, the packages are relatively small compared to the respective Raspbian packages, based on Debian and designed for desktop class computers.
For the \texttt{wget} examples, the downloaded data is not written to disk, and thus only shows the performance of the networking stack. 
Downloading the file using HTTP shows the superiority of the build server, only requiring a tenth of the time compared to native execution.
Since most build servers do have larger than 1 GBit/s Internet connections, the runtimes can be up to 10 times as fast compared to a Raspberry Pi limited to 100 MBit/s.
However, when using the same protocol with TLS encryption, the cost of encryption and decryption adds an overhead of 13.6\%.
In the examples using \texttt{dd}, 100 MB of random data is read using the command \texttt{dd if=/dev/urandom bs=1M count=100}. 
The computations needed to generate the data happen in the kernel and thus natively on both systems.
When discarding the data using \texttt{of=/dev/null}, no heavy computation is required and \pimod performs better, because the main load happens in the kernel of the host.
From the runtimes of the second example presented in the table it becomes evident that for read/write intensive tasks, the disk is the bottleneck; the overhead gain of 91.24\% is comparable to the download task. 
In the last example, data is read from \texttt{/dev/urandom} and encrypted using the OpenSSL AES encryption. 
The speed advantage of \pimod fades, since the computationally intensive encryption is performed through the QEMU emulation layer. 
Nevertheless, our approach is still around 30\% faster compared to native execution.

\subsection{Testing Linux Distributions}
To test the generalization properties of our approach, two directions are important: hardware and operating system variety.
In our tests using the well known Raspberry Pi, we were able to use \pimod for the widespread Debian-based operating systems, such as Raspbian and Ubuntu Server, as well as OpenWRT, CentOS, Fedora, Kali, and OpenSUSE.
Android as well as NixOS did not implement the Filesystem Hierarchy Standard and therefore do not satisfy our assumptions.
\pimod can be adapted to work with both distributions, but a generalization is not easily feasible.
In addition, we evaluated different hardware platforms using the operating system recommended and distributed by the vendor.
In our tests, we found that \pimod can be used with the Libre Computer boards ALL-H3-CC, AML-S805X-AC and ROC-RK3328-CC, BananaPi M4, Nvidia Jetson Nano (AI development board), ODROID C2 and N2, OrangePi 3, all models of RaspberryPi, and the RockPi 4.
Most of the operating systems distributed by the vendors are Debian-based and include specific changes for the individual boards.

\subsection{\pimod Language Flexibility}
One secondary design goal was to support flexibility for the users of \pimod. 
The presented tool is written in the Bash scripting language, a Pifile itself is a script sourced in the individual stages.
Thus, in a Pifile all features available in Bash scripts can be used.
\pimod's flexibility features will be discussed by the examples presented in Listing~\ref{lst:example2}, the respective output log is shown in Listing~\ref{lst:example2log}.

\begin{lstlisting}[
    style=Pimod, 
    caption={\pimod example 2: advanced scripting with Bash features.}, 
    label={lst:example2}, 
    float=ht,
]
FROM http://downloads.openwrt.org/releases/18.06.5/targets/brcm2708/bcm2710/openwrt-18.06.5-brcm2708-bcm2710-rpi-3-ext4-factory.img.gz

# Derive block device from environment
TO $DEVICE

# Include wifi configuration 
source modules/wifi.Pifile

# Add local public ssh key
RUN tee -a /etc/dropbear/authorized_keys <$HOME/.ssh/id_rsa.pub

# Set DHCP client mode for eth0
RUN tee -a /etc/config/network <<EOF
config interface 'lan'
        option type 'bridge'
        option ifname 'eth0'
        option proto 'dhcp'
EOF

# Cross-compile local software
HOST GOOS=linux GOARCH=arm GOARM=5 go build -o dtn7d ./dtn7-go/cmd/dtnd
INSTALL 755 dtn7d /usr/bin/dtn7d
\end{lstlisting}

\begin{lstlisting}[
    style=console, 
    caption={\pimod example 2 (Listing~\ref{lst:example2}) execution log.}, 
    label={lst:example2log}, 
    float=ht,
]
### FROM http://downloads.openwrt.org/releases/18.06.5/targets/brcm2708/bcm2710/openwrt-18.06.5-brcm2708-bcm2710-rpi-3-ext4-factory.img.gz
Using cache: /var/cache/pimod/downloads.openwrt.org/releases/18.06.5/targets/brcm2708/bcm2710/openwrt-18.06.5-brcm2708-bcm2710-rpi-3-ext4-factory.img.gz
### TO sd.img
Moving temporary /tmp/tmp.sv3nxlM9qS to sd.img
add map loop0p1 (253:0): 0 40960 linear 7:0 8192
add map loop0p2 (253:1): 0 524288 linear 7:0 57344
### RUN tee -a /etc/dropbear/authorized_keys
ssh-rsa AAAA.... hoechst@ds
### RUN tee -a /etc/config/network
config interface 'lan'
        option type 'bridge'
        option ifname 'eth0'
        option proto 'dhcp'
### HOST pushd dtn7-go
/storage/hoechst/pimod-example/dtn7-go /storage/hoechst/pimod-example
### HOST go build -o dtn7d ./cmd/dtnd
### INSTALL 755 dtn7d /usr/bin/dtn7d
### HOST popd
/storage/hoechst/pimod-example
umount: /tmp/tmp.ampfPoR8rz/dev/pts unmounted
umount: /tmp/tmp.ampfPoR8rz/dev unmounted
umount: /tmp/tmp.ampfPoR8rz/sys unmounted
umount: /tmp/tmp.ampfPoR8rz/proc unmounted
umount: /tmp/tmp.ampfPoR8rz unmounted
del devmap : loop0p2
del devmap : loop0p1
\end{lstlisting}

\subsubsection{Environment Variables}
Environment variables can be helpful to adjust the build to the runtime. 
In the example presented in Line 5 of Listing~\ref{lst:example2}, a block device to work on can be specified through the environment variable \texttt{DEVICE}. 
In Line 11, another example is shown, where the ssh public key from the local user is added to the guest system.

\subsubsection{Redirections}
Redirections of input and output streams are a key feature in shell programming and can be helpful in our use cases.
The task of adding an ssh key from the host system, presented in Line 11, is implemented by appending to a file and redirecting the input stream to a file in the host system.
The \texttt{HOME} variable provides the location of the host system user's home directory. 
A second example is presented in Lines~14--19, where a network configuration file is written by using a \textit{here document}. 
When using documents, the configuration resides inside of the Pifile itself, and is easily understandable. 
Environment variables are also evaluated inside of \textit{here documents}, such that device-specific configuration, e.g., a hostname, can be set using this mechanism.

\subsubsection{Cross-Compilation}
For some cases, it is beneficial to compile software on the host system, rather than running the virtualized compiler of the guest system, especially for larger software projects.
In the above example of Line~22, the delay-tolerant routing software DTN7 \cite{penning2019dtn7} is cross-compiled on the host system by using the \texttt{HOST} command and installed to the guest system afterwards.

\subsubsection{Modularization}
Individual parts of a Pifile can be extracted into modules, represented as independent files, that can be reused afterwards. 
In the example presented above, the Wi-Fi configuration is included from another Pifile, as presented in Line~8.
The file is included by using the \texttt{source} command, which executes the commands from the file in the current script.
Thus, \pimod offers a wide variety of interactions with the host systems and can thereby be integrated into user's workflows well.

\subsection{Applications of \pimod}

\begin{figure}[tb]
    \centering
    \vspace{.05in}
    \includegraphics[width=\columnwidth]{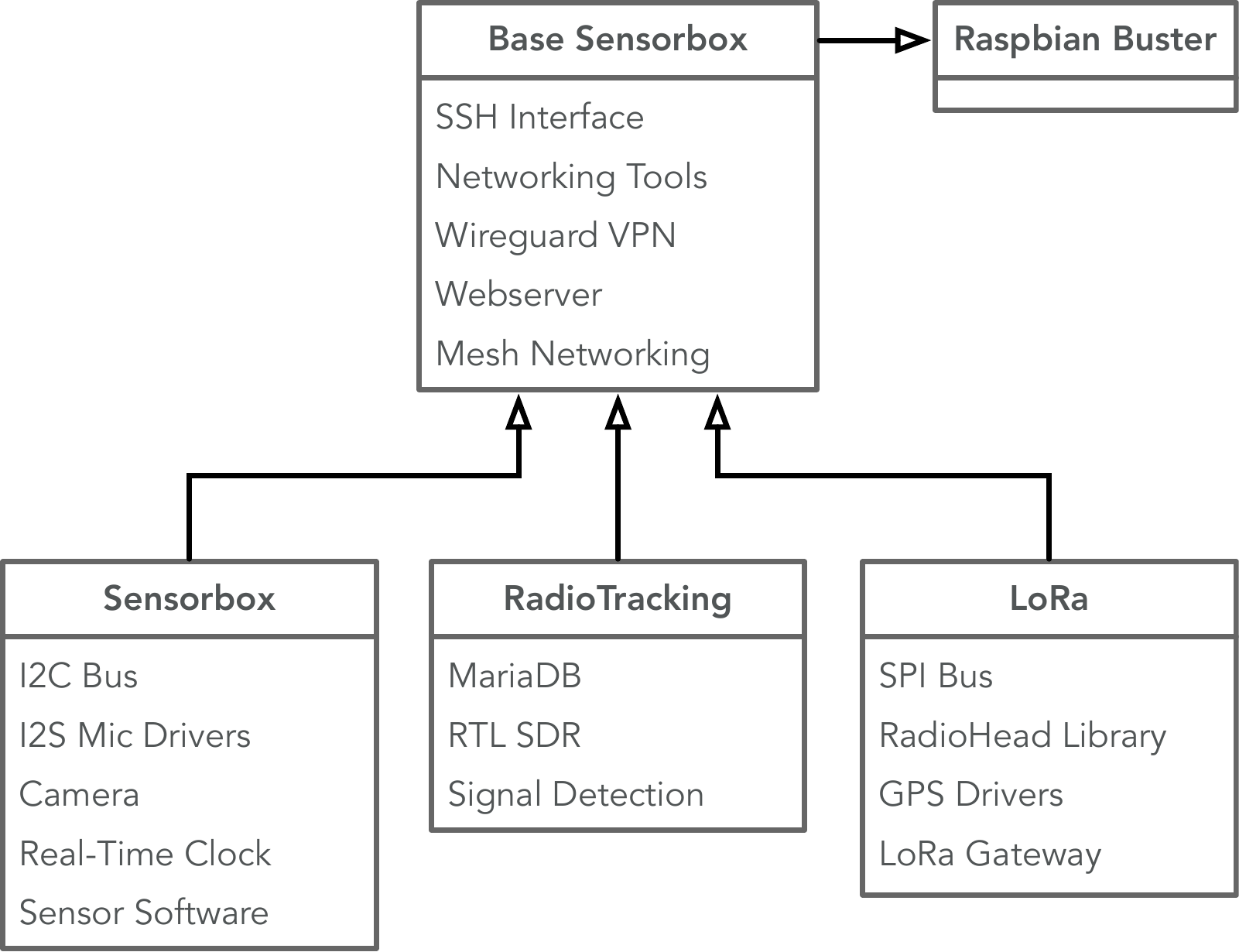}
    \caption{Raspberry Pi Image configurations used for the Nature 4.0 Project}
    \label{figure:nature40_sensorboxes}
\end{figure}

The goal of the Nature 4.0 project is to develop a modular environmental monitoring system for high-resolution observation of species, habitats, and processes relevant to nature conservation \cite{friess2019introducing}. 
For this use case, a set of sensor boxes has been designed, each dedicated to sightly different use cases, following the same principal networking functionalities and the same software basis, each based on a Raspberry Pi.
In Fig.~\ref{figure:nature40_sensorboxes}, the hierarchy derived from the requirements of the project are presented.
The first box is the rather generic \textit{Sensorbox}, which comes in different configurations, distinguished by the wired sensor set. 
It includes a camera, a microphone via the I2S bus, sensors for temperature, humidity, atmospheric pressure, air quality, spectral brightness, and various others.
The second box is the \textit{RadioTracking} configuration. 
This box is dedicated to read and analyze data flows of four software-defined radio devices per box, which point to the four directions of the compass in order to record signals from bats tagged with special senders \cite{gottwald2019introduction}.
The last configuration is a \textit{LoRa} gateway receiving and forwarding messages of other sensor applications via the well known long range \textit{LoRa} protocol. 
All images are derived from the \textit{Raspbian} OS, in its \textit{Buster} version, which is the 
de facto standard for Raspberry Pi SBCs.
The \textit{Base Sensorbox} image contains all shared software and configuration, e.g., an access option via a VPN and a SSH remote shell, a web server for convenient data retrieval and management in the field, as well as a mesh network configuration for sensor box interconnection.

\pimod is used in the project Nature 4.0 for over a year at the date of writing, and integrates well into the workflow\footnote{\url{https://github.com/nature40/sensorboxes-images/}}.
A software or system developer can get a copy of the project using the git versioning system, including all sub modules, and can build am image from scratch in the matter of minutes. 
Additional software as well as adjustments can be implemented and a new image can be built and tested locally.
After a successful test, the changes are committed, and a new release is built and using the CI integration mentioned above.

%% file: 6_conclusion.tex
\section{Conclusion}
\label{sec:conclusion}

In this paper. we presented \pimod, a tool for configuring single-board computer operating system images. 
A simple yet comprehensive language for configuring these images was proposed, which offers an interface to operating system specific configuration tools.
The implementation of the proposed language was presented and the staged execution of a Pifile was explained.
\pimod was evaluated by first discussing the benefits for operators of a sensor network.
The performance of the proposed approach was evaluated by comparing runtimes of exemplary commands providing an overview of the overhead.
Furthermore, it was demonstrated that \pimod supports a wide variety of hardware platforms and operating systems.

An interesting area of future work is to extend \pimod to enable reproducible image builds.
However, some of the unique characteristics of \pimod's implementation prevent hash identical builds. 
Several commands executed in the guest systems have non-deterministic side effects, such as creating temporary and logging files.
Another example are the time stamps contained in the image file system, which would also need to be adjusted accordingly.

\section{Acknowledgements}
This work is funded by the Hessian State Ministry for Higher Education, Research and the Arts (HMWK) (LOEWE Natur 4.0 and LOEWE emergenCITY), and the Deutsche Forschungsgemeinschaft (DFG, SFB 1053).

%% file: main.bbl
\begin{thebibliography}{10}
\providecommand{\url}[1]{#1}
\csname url@samestyle\endcsname
\providecommand{\newblock}{\relax}
\providecommand{\bibinfo}[2]{#2}
\providecommand{\BIBentrySTDinterwordspacing}{\spaceskip=0pt\relax}
\providecommand{\BIBentryALTinterwordstretchfactor}{4}
\providecommand{\BIBentryALTinterwordspacing}{\spaceskip=\fontdimen2\font plus
\BIBentryALTinterwordstretchfactor\fontdimen3\font minus
  \fontdimen4\font\relax}
\providecommand{\BIBforeignlanguage}[2]{{%
\expandafter\ifx\csname l@#1\endcsname\relax
\typeout{** WARNING: IEEEtranS.bst: No hyphenation pattern has been}%
\typeout{** loaded for the language `#1'. Using the pattern for}%
\typeout{** the default language instead.}%
\else
\language=\csname l@#1\endcsname
\fi
#2}}
\providecommand{\BIBdecl}{\relax}
\BIBdecl

\bibitem{baumgartner2018environmental}
L.~Baumg{\"a}rtner, A.~Penning, P.~Lampe, B.~Richerzhagen, R.~Steinmetz, and
  B.~Freisleben, ``Environmental monitoring using low-cost hardware and
  infrastructureless wireless communication,'' in \emph{2018 IEEE Global
  Humanitarian Technology Conference (GHTC)}.\hskip 1em plus 0.5em minus
  0.4em\relax IEEE, 2018, pp. 1--8.

\bibitem{baumgaertner2019smart}
L.~{Baumgärtner}, J.~{Höchst}, P.~{Lampe}, R.~{Mogk}, A.~{Sterz},
  P.~{Weisenburger}, M.~{Mezini}, and B.~{Freisleben}, ``{Smart street lights
  and mobile citizen apps for resilient communication in a digital city},'' in
  \emph{2019 IEEE Global Humanitarian Technology Conference (GHTC)}, 2019, pp.
  1--8.

\bibitem{baun2016mobile}
C.~Baun, ``{Mobile clusters of single board computers: an option for providing
  resources to student projects and researchers},'' \emph{SpringerPlus},
  vol.~5, no.~1, p. 360, 2016.

\bibitem{bellard2005qemu}
F.~Bellard, ``{QEMU, a fast and portable dynamic translator.}'' in \emph{USENIX
  Annual Technical Conference, FREENIX Track}, vol.~41, 2005, p.~46.

\bibitem{beller2017oops}
M.~Beller, G.~Gousios, and A.~Zaidman, ``{Oops, my tests broke the build: An
  explorative analysis of Travis CI with GitHub},'' in \emph{2017 IEEE/ACM 14th
  International Conference on Mining Software Repositories (MSR)}.\hskip 1em
  plus 0.5em minus 0.4em\relax IEEE, 2017, pp. 356--367.

\bibitem{pishrink}
\BIBentryALTinterwordspacing
D.~Bonasera. (2016) {PiShrink: Make your Pi images smaller!} [Online].
  Available: \url{https://github.com/Drewsif/PiShrink}
\BIBentrySTDinterwordspacing

\bibitem{daigle2018github}
K.~Daigle. (2018, October) {GitHub Actions: built by you, run by us}. GitHub.

\bibitem{fainelli2008openwrt}
F.~Fainelli, ``{The OpenWRT embedded development framework},'' in
  \emph{Proceedings of the Free and Open Source Software Developers European
  Meeting}, 2008, p. 106.

\bibitem{PiBakery}
\BIBentryALTinterwordspacing
D.~Ferguson. (2016) {PiBakery: Easily customise Raspbian}. [Online]. Available:
  \url{https://www.pibakery.org/index.html}
\BIBentrySTDinterwordspacing

\bibitem{friess2019introducing}
N.~Friess, J.~Bendix, M.~Br{\"a}ndle, R.~Brandl, S.~Dahlke, N.~Farwig,
  B.~Freisleben, H.~Holzmann, H.~Meyer, T.~M{\"u}ller \emph{et~al.},
  ``{Introducing Nature 4.0: A sensor network for environmental monitoring in
  the Marburg Open Forest},'' \emph{Biodiversity Information Science and
  Standards}, 2019.

\bibitem{gottwald2019introduction}
J.~Gottwald, R.~Zeidler, N.~Friess, M.~Ludwig, C.~Reudenbach, and T.~Nauss,
  ``Introduction of an automatic and open-source radio-tracking system for
  small animals,'' \emph{Methods in Ecology and Evolution}, vol.~10, no.~12,
  pp. 2163--2172, 2019.

\bibitem{hochstein2017ansible}
L.~Hochstein and R.~Moser, \emph{{Ansible: Up and Running: Automating
  configuration management and deployment the easy way}}.\hskip 1em plus 0.5em
  minus 0.4em\relax {O'Reilly Media, Inc.}, 2017.

\bibitem{hosmer2012getting}
B.~Hosmer, ``{Getting started with Salt Stack - the other configuration
  management system built with Python},'' \emph{Linux journal}, vol. 2012, no.
  223, p.~3, 2012.

\bibitem{johnston2018commodity}
S.~J. Johnston, P.~J. Basford, C.~S. Perkins, H.~Herry, F.~P. Tso, D.~Pezaros,
  R.~D. Mullins, E.~Yoneki, S.~J. Cox, and J.~Singer, ``{Commodity single board
  computer clusters and their applications},'' \emph{Future Generation Computer
  Systems}, vol.~89, pp. 201--212, 2018.

\bibitem{Kairiukstis2012BuildRaspbianImage}
\BIBentryALTinterwordspacing
A.~Kairiukstis. (2012) {BuildRaspbianImage: Build (and cross-compile) your own
  image for Raspberry Pi}. [Online]. Available:
  \url{https://github.com/andrius/build-raspbian-image/}
\BIBentrySTDinterwordspacing

\bibitem{kumar2016iot}
R.~Kumar and M.~P. Rajasekaran, ``{An IoT based patient monitoring system using
  Raspberry Pi},'' in \emph{2016 International Conference on Computing
  Technologies and Intelligent Data Engineering (ICCTIDE'16)}.\hskip 1em plus
  0.5em minus 0.4em\relax IEEE, 2016, pp. 1--4.

\bibitem{loope2011managing}
J.~Loope, \emph{{Managing infrastructure with puppet: Configuration management
  at scale}}.\hskip 1em plus 0.5em minus 0.4em\relax {O'Reilly Media, Inc.},
  2011.

\bibitem{merkel2014docker}
D.~Merkel, ``{Docker: Lightweight Linux containers for consistent development
  and deployment},'' \emph{Linux Journal}, vol. 2014, no. 239, p.~2, 2014.

\bibitem{penning2019dtn7}
A.~Penning, L.~Baumgärtner, J.~Höchst, A.~Sterz, M.~Mezini, and
  B.~Freisleben, ``{DTN7: An open-Source disruption-tolerant networking
  implementation of Bundle Protocol 7},'' in \emph{International Conference on
  Ad-Hoc Networks and Wireless (AdHoc-Now 2019)}.\hskip 1em plus 0.5em minus
  0.4em\relax Luxembourg, Luxembourg: Springer, 2019, pp. 196--209.

\bibitem{quitevis2018feasibility}
C.~P. Quitevis and C.~D. Ambatali, ``{Feasibility of an amateur radio
  transmitter implementation using Raspberry Pi for a low cost and portable
  emergency communications device},'' in \emph{2018 IEEE Global Humanitarian
  Technology Conference (GHTC)}.\hskip 1em plus 0.5em minus 0.4em\relax IEEE,
  2018, pp. 1--6.

\bibitem{ramey1994s}
C.~Ramey, ``{What's GNU: Bash - The GNU Shell},'' \emph{Linux Journal}, vol.
  1994, no. 4es, p.~13, 1994.

\bibitem{raspbian}
\BIBentryALTinterwordspacing
{Raspberry Pi Foundation}. (2014) {Raspbian: A free operating system based on
  Debian optimized for the Raspberry Pi hardware}. [Online]. Available:
  \url{https://www.raspberrypi.org/downloads/raspbian/}
\BIBentrySTDinterwordspacing

\bibitem{pigen}
\BIBentryALTinterwordspacing
------. (2016) {pi-gen: Tool used to create the raspberrypi.org Raspbian
  images}. [Online]. Available: \url{https://github.com/RPi-Distro/pi-gen}
\BIBentrySTDinterwordspacing

\bibitem{srinivasan2013greeneducomp}
M.~Srinivasan, A.~V. AJ, A.~N. Victor, M.~Narayanan, S.~R. SP,
  V.~Vijayaraghavan \emph{et~al.}, ``{GreenEduComp: Low cost green computing
  system for education in Rural India: A scheme for sustainable development
  through education},'' in \emph{2013 IEEE Global Humanitarian Technology
  Conference (GHTC)}.\hskip 1em plus 0.5em minus 0.4em\relax IEEE, 2013, pp.
  102--107.

\bibitem{truitt2019low}
S.~{Truitt}, T.~D. {Gage}, B.~E. {Vincent}, and S.~{Chun}, ``{Low-cost remote
  monitoring system for small-Scale UPS installations in developing
  countries},'' in \emph{2019 IEEE Global Humanitarian Technology Conference
  (GHTC)}, 2019, pp. 1--6.

\bibitem{vasilescu2015quality}
B.~Vasilescu, Y.~Yu, H.~Wang, P.~Devanbu, and V.~Filkov, ``{Quality and
  productivity outcomes relating to continuous integration in GitHub},'' in
  \emph{Proceedings of the 2015 10th Joint Meeting on Foundations of Software
  Engineering}, 2015, pp. 805--816.

\bibitem{yamanoor2017high}
N.~S. Yamanoor and S.~Yamanoor, ``{High quality, low cost education with the
  Raspberry Pi},'' in \emph{2017 IEEE Global Humanitarian Technology Conference
  (GHTC)}.\hskip 1em plus 0.5em minus 0.4em\relax IEEE, 2017, pp. 1--5.

\end{thebibliography}
